\documentclass[
aip,
amsmath,
amssymb,
reprint
]{revtex4-1}

\usepackage{graphicx}
\usepackage{dcolumn}
\usepackage{bm}

\usepackage[utf8]{inputenc}
\usepackage[T1]{fontenc}
\usepackage{mathptmx}
\usepackage{etoolbox}
\usepackage{amsmath}

\usepackage{parskip}

\usepackage{hyperref}
\usepackage{xcolor}

\usepackage{placeins}
\usepackage{float}

\hypersetup{
    colorlinks=false,     
    linkbordercolor=green,
    urlbordercolor=green, 
    pdfborderstyle={/S/U/W 1} 
}

\makeatletter
\def\@email#1#2{%
 \endgroup
 \patchcmd{\titleblock@produce}
  {\frontmatter@RRAPformat}
  {\frontmatter@RRAPformat{\produce@RRAP{*#1\href{mailto:#2}{#2}}}\frontmatter@RRAPformat}
  {}{}
}%
\makeatletter

\begin{document}

\preprint{AIP/123-QED}

\title{Engineering wavefronts with machine learned structured polarization}

\author{Sai Nikhilesh Murty Kottapalli}
    \email{nikhilesh.kottapalli@mr.mpg.de}
    \affiliation{Max Planck Institute for Medical Research, 69120 Heidelberg, Germany}
    \affiliation{Institute for Molecular Systems Engineering and Advanced Materials, Universität Heidelberg, 69120 Heidelberg, Germany}

\author{Alexander Song}%
    \email{alexander.song@mr.mpg.de}
    \affiliation{Max Planck Institute for Medical Research, 69120 Heidelberg, Germany}
    \affiliation{Institute for Molecular Systems Engineering and Advanced Materials, Universität Heidelberg, 69120 Heidelberg, Germany}%

\author{Peer Fischer}
    \affiliation{Max Planck Institute for Medical Research, 69120 Heidelberg, Germany}
    \affiliation{Institute for Molecular Systems Engineering and Advanced Materials, Universität Heidelberg, 69120 Heidelberg, Germany}
    \affiliation{Center for Nanomedicine, Institute for Basic Science (IBS), Seoul, Republic of Korea}
    \affiliation{Department of Nano Biomedical Engineering (NanoBME), Advanced Science Institute, Yonsei University, Seoul, Republic of Korea}

\begin{abstract}
  Optical approaches for wavefront shaping traditionally rely on phase modulation through holographic techniques. Shaping the phase determines a wave's diffraction and hence its intensity distribution in space.
  We instead show that shaping the polarization introduces a novel framework that permits the spatial modulation of polarization to control wavefront propagation and resulting amplitude distributions.  
  We develop two distinct computational phase retrieval approaches for calculating the required polarization transformations and experimentally validate these. The first method extends the established Gerchberg-Saxton algorithm, while the second employs machine learning optimization to determine optimal polarization patterns. By implementing both amplitude and polarization control simultaneously using a single polarization mask, our approach significantly reduces system complexity compared to traditional methods. Our experimental results demonstrate the potential of polarization-based wavefront shaping as an efficient alternative to conventional techniques, paving the way for applications in optical manipulation and imaging.
\end{abstract}

\maketitle

\section{Introduction}
\label{sec: introduction}

The wave nature of light gives rise to diffraction and interference phenomena, forming the foundation of modern optical systems. While these phenomena are widely exploited in holography for wavefront shaping, existing approaches predominantly rely on phase modulation to achieve desired intensity distributions \cite{hariharan1996optical}. Recent advances in optical technologies have created new opportunities for wavefront control using alternative approaches. This study introduces a novel framework that utilizes spatially varying polarization modulation, rather than phase modulation, to shape propagating wavefronts and control the resulting amplitude distributions.

Research in optical wavefront shaping has been explored for decades \cite{dickey2018laser} and is currently undergoing significant advancement. The combination of improved spatial light modulators (SLMs) and digital micromirror devices (DMDs) with sophisticated control algorithms has led to promising results in beam control and shaping.  However, these approaches primarily operate through spatial modulation of either phase or amplitude, while the potential of polarization-based modulation remains largely unexplored.

Several challenges need to be addressed before polarization-based wavefront shaping and beam steering can compete with existing phase modulation techniques.  Current methods often rely on the global rotation of polarization \cite{fratz2009digital} and they generally achieve only limited beam displacements \cite{salazar2015demonstration}. While some techniques have successfully generated specific beam types, such as vortex beams \cite{dennis2009singular}, or achieved flat-top focusing \cite{chen2014demonstration}, they typically do not permit dynamic reconfigurability \cite{brown2013stress}, limiting their practical applications.

The challenge for polarization-based wavefront shaping lies in developing a general framework capable of optimizing polarization modulation to allow arbitrary changes to the resulting intensity distribution. Standard phase-retrieval algorithms for holography utilize computationally intensive physics-based models to calculate the requisite phase modulation.  However, these algorithms are not well-suited for calculating polarization modulation. Data-driven approaches, such as machine learning, have shown promise in optimizing phase masks for wavefront shaping \cite{rivenson2018phase}.  The use of machine learning optimization techniques has opened up new possibilities, such as the ability to optimize for multiple target patterns simultaneously \cite{zhang20173d}, thereby opening up various applications \cite{melde2023compact}.

In this work, we demonstrate the potential of non-uniform spatial polarization structures for wavefront control. We introduce and experimentally demonstrate a framework based on polarization masks that can generate arbitrary intensity patterns. Our system builds upon and extends prior work implementing polarization singularities \cite{dennis2009singular} and beam shaping techniques.  We develop two distinct computational approaches for calculating the required polarization transformations: a modified Gerchberg-Saxton algorithm and a machine learning optimization technique.

\begin{figure*}[bt]
    \centering
    \includegraphics[width=0.95\textwidth]{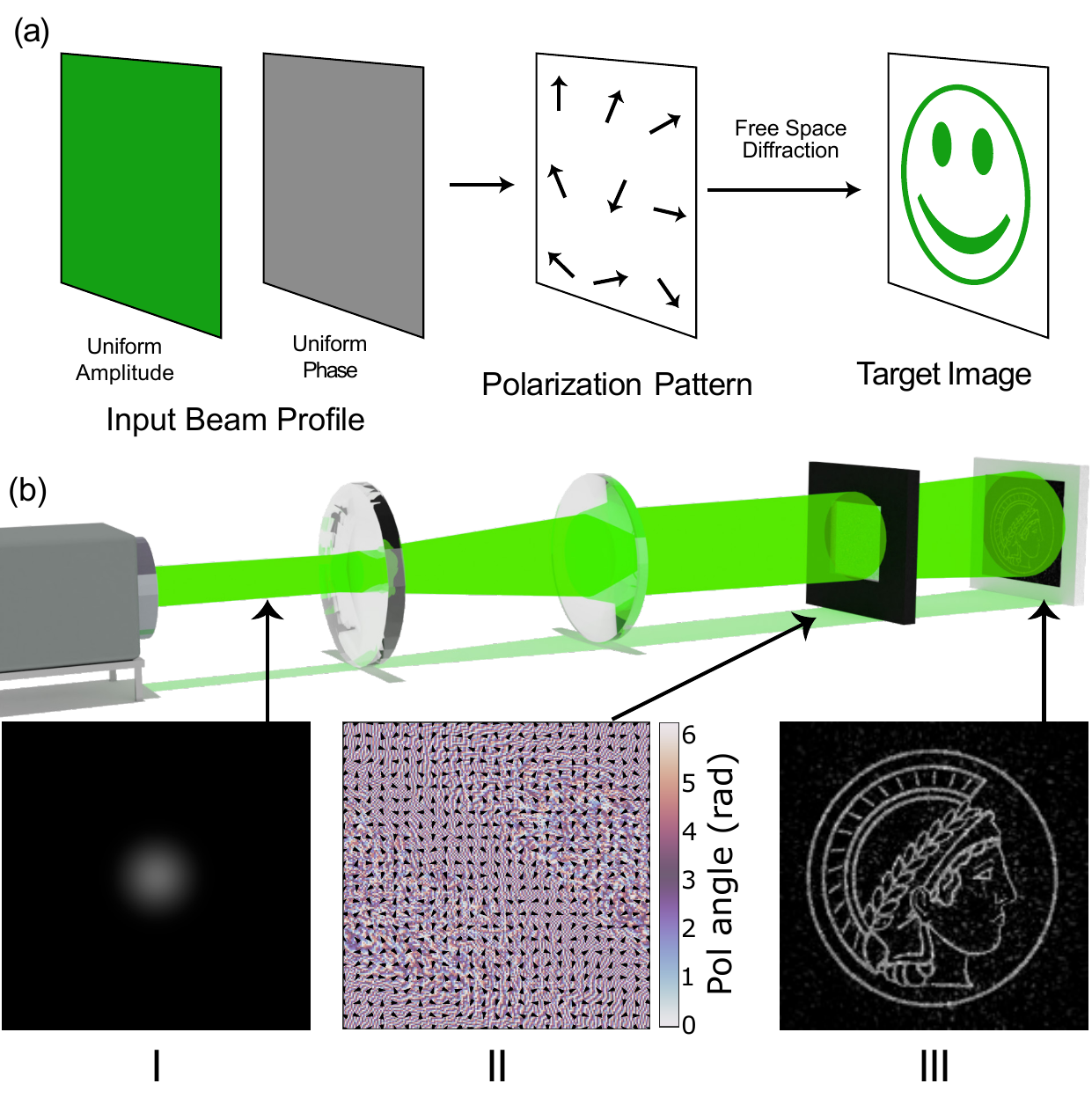}
    \caption{Schematic of the experimental setup. (a) Conceptual overview: a polarization modulation mask transforms an input wavefront with a uniform (or Gaussian) amplitude and phase into a desired target amplitude and polarization distribution ("Target Image") at the target plane. (b) Experimental implementation: (I) A Gaussian laser beam is expanded and then spatially modulated by a twisted nematic liquid crystal spatial light modulator (SLM). (II) The SLM imparts a spatially inhomogeneous polarization rotation to the linearly polarized input light. (III) This polarization modulation results in the desired intensity distribution at the target plane, which is recorded by a camera.}
    \label{fig:schematic}
\end{figure*}

Our work is experimentally realized using a coherent light source and a twisted nematic liquid crystal SLM (FIG. \ref{fig:schematic}).  The focus of this work is to demonstrate a polarization-based wavefront shaping paradigm that can be straightforwardly implemented for a variety of target intensity patterns. The system is designed so that both amplitude and polarization can be controlled simultaneously at the target plane using a single polarization mask, a capability not previously demonstrated.

These advantages, coupled with the ability to dynamically reconfigure the polarization masks, pave the way for new applications in microscopy, lithography, and optical manipulation. Our approach is general and extensible to various types of target patterns while maintaining the ability to operate in real-time, thus suggesting broad applicability in fields requiring precise optical control.

\section{Methods}
\label{sec:methods}
This work focuses on optimizing the wavefront at a target plane via polarization modulation of light at the object plane. The experimental configuration is depicted in FIG. \ref{fig:schematic}. A 532 nm Gaussian laser source is expanded with a beam expander and propagates through a half-wave plate that rotates the beam's polarization to orient it with the liquid crystal director of a spatial light modulator (SLM, Holoeye LC2012). Subsequently, the beam interacts with the SLM, a twisted nematic SLM where the twist angle is electrically adjustable. This variation in twist angle rotates the polarization of the transmitted beam and the $1024 \times 768$ pixelated electrode surface functions as a locally variable wave-plate with the same resolution. Although this process introduces a geometric phase, this contribution has a negligible effect on the amplitude distribution at the target plane (supplementary section 1), which is the primary focus of our measurements. Finally, the beam propagates to the target plane, where a camera (Flir ORX-10G-71S7M-C) detects the intensity.

To obtain a desired target amplitude distribution, it is necessary to compute the corresponding polarization mask that is applied to the SLM. The angular spectrum method (ASM) \cite{goodman2005introduction} models the propagation of an optical field from the object plane to the target plane and is an efficient computational method for simulating the propagation of optical fields in free space. With the ASM, the initial complex optical field $U_s(x,y,0)$ at the object plane ($z=0$) is decomposed into plane waves through a 2D Fourier transform:

\begin{equation}
    A(k_x, k_y;0) = \iint_{-\infty}^{\infty} U_s(x,y,0) e^{-i(k_xx+k_yy)}dxdy
    \label{eq:asm_angular_spectrum}
\end{equation}

where the angular spectrum of the field is given by $A(k_x, k_y;0)$, with $k_x$ and $k_y$ as the spatial frequencies along the $x$ and $y$ directions, respectively. These spatial frequencies correlate with the direction cosines of the plane waves. The field at the target plane (distance $z$) is reconstructed by accounting for the phase accumulation of each plane wave component during propagation:

\begin{equation}
    U(x,y,z) = \iint_{-\infty}^{\infty} A(k_x, k_y;0) e^{i(k_xx+k_yy+k_zz)}dk_xdk_y
    \label{eq:asm_angular_spectrum_reconstruction}
\end{equation}

Here, $k_z = \sqrt{k^2 - k_x^2 - k_y^2}$ gives the longitudinal component of the wavevector, and $k = \frac{2\pi}{\lambda}$ is the wavenumber, and $\lambda$ is the wavelength. If $k^2 < k_x^2 + k_y^2$, then $k_z$ becomes imaginary, representing evanescent waves. These waves decay exponentially and they do not propagate to the far field. The above two equations can be described as a composition of functions, stated as follows:

\begin{equation}
    U(x,y,z) = \mathcal{F}^{-1}\{\mathcal{F}\{U(x,y,0)\}H(k_x,k_y;z)\}
    \label{eq:asm_composition}
\end{equation}

where $\mathcal{F}$ and $\mathcal{F}^{-1}$ denote the Fourier transform and inverse Fourier transform, respectively, and $H(k_x,k_y;z)$ is the transfer function describing free-space propagation.

This model captures the wave diffraction. To incorporate the polarization, the light's polarization is described using the Jones vector, a two-element complex vector:

\begin{equation}
    \begin{pmatrix}
    E_x \\
    E_y
    \end{pmatrix}
    \label{eq:jones_vector}
\end{equation}

Each element of this vector gives the amplitude and phase of the electric field in two orthogonal states. These basis states, linear in our case, are treated independently as they do not interact during propagation.

Since the SLM operates as a pixel-by-pixel rotator, its effect on light polarization is described as a rotation matrix acting on the Jones vector. A rotation by $\theta$ is represented by the matrix:

\begin{equation}
    R(\theta) = \begin{pmatrix}
    \cos(\theta) & -\sin(\theta) \\
    \sin(\theta) & \cos(\theta)
    \end{pmatrix}
    \label{eq:jones_rotation_matrix}
\end{equation}

These equations are used as the forward and backward operators for both algorithmic approaches for estimating the applied polarization mask.

\section{Results}
\label{sec:results}

This work demonstrates the use of spatially varying polarization modulation to control wavefront propagation for achieving desired wavefronts after propagation.  As a proof of concept experiment we first show that a suitable polarization mask can convert a Gaussian beam into a focused pseudo-Bessel beam at the target plane (see Supplementary Section \ref{si:focussing} for further details). Here, we discuss how polarization masks can be used to generate arbitrary 2d intensity patterns in the target plane. Two distinct computational approaches are developed for calculating the required polarization transformations, and the results obtained with these methods are demonstrated experimentally. The first method extends the established Gerchberg-Saxton algorithm, whereas the second employs machine learning optimization to determine optimal polarization mask patterns.  Furthermore, we demonstrate the joint optimization of both phase and polarization using machine learning techniques. The results of these experiments are presented in the following subsections.

\subsection{Conventional Holographic Techniques}
\label{subsec:traditional_methods}

The central challenge lies in calculating the polarization modulation of the wavefront that, upon diffraction, produces the desired amplitude distribution at the target plane. A commonly employed approach to solve this type of problem involves iterative phase retrieval algorithms. We adapt a widely-used phase retrieval technique, the Gerchberg-Saxton algorithm \cite{yang1994gerchberg}, to our experimental setup. This adaptation enables us to iteratively refine the polarization mask and achieve the target amplitude pattern.

The iterative optimization procedure starts with the field at the object plane, denoted as $U_s$. $U_s$ has a two-dimensional Gaussian amplitude profile, $A_s$, and a uniform (flat) phase front. A random phase distribution is applied to $U_s$ as an initial guess.   This field is then numerically propagated to the target plane using the angular spectrum method (ASM), as detailed in Section \ref{sec:methods}.

At the target plane, located at a distance  $z$ from the object plane, the propagated field, $U_p$, is characterized by an amplitude $A_{t'}$ and phase $\phi_{t'}$. The objective of the Gerchberg-Saxton algorithm is to iteratively refine the phase at the object plane (and consequently, indirectly determine the required polarization mask) such that the amplitude of the propagated field closely approximates the desired target amplitude $A_t$. The individual steps of this modified Gerchberg-Saxton algorithm are summarized visually in Fig. \ref{fig:traditional_holography}a and elaborated upon below:

\begin{figure*}[!hbt]
    \centering
    \includegraphics[width=0.95\textwidth]{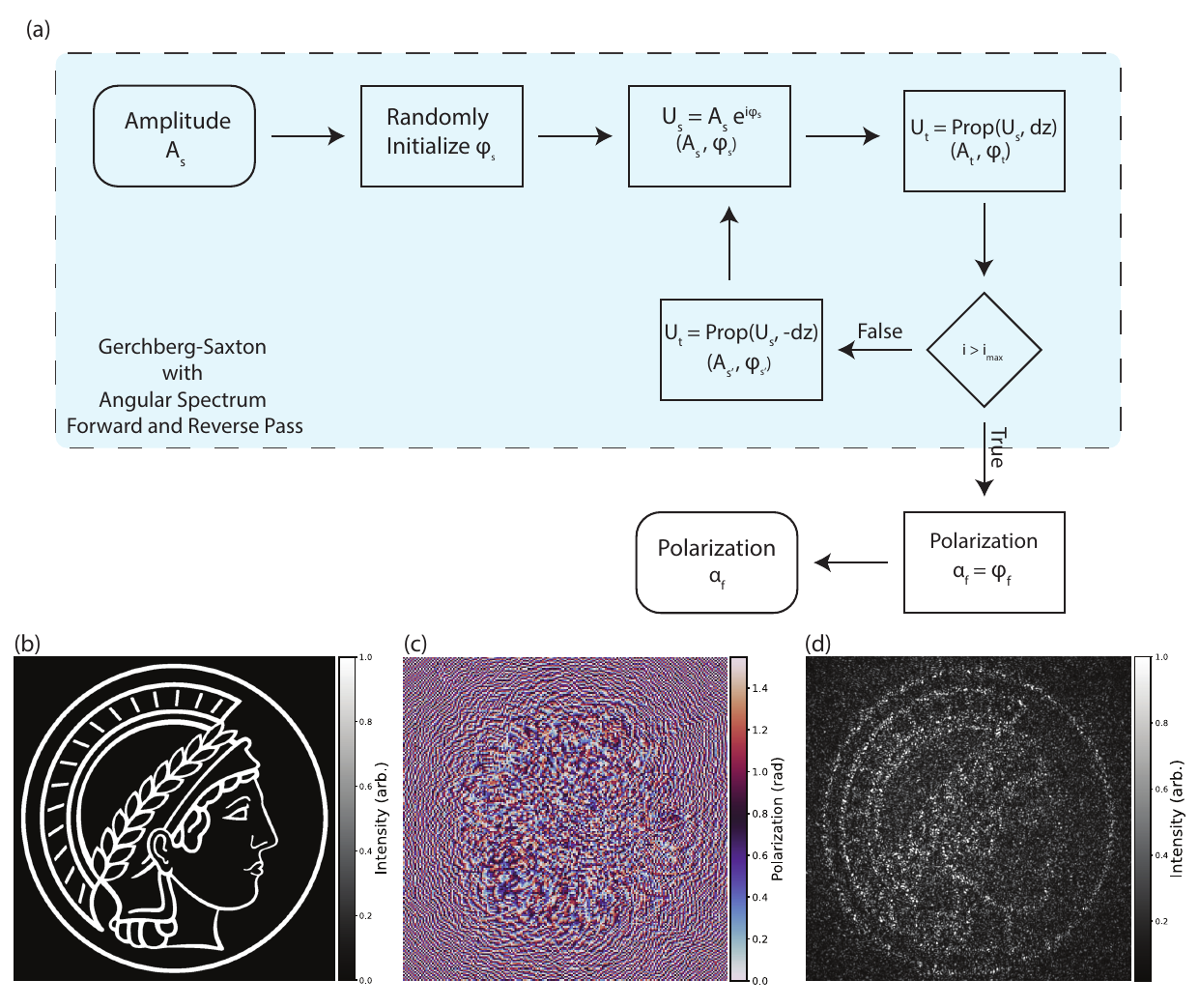}
    \caption{Optimization of a polarization mask using a modified Gerchberg-Saxton algorithm. The figure is divided into two parts. The first part, highlighted with a blue background, illustrates the conventional Gerchberg-Saxton method for phase retrieval.  An initial random phase mask $\phi_s$ is defined, and the input amplitude $A_s$ is propagated to the target plane using the angular spectrum method.  An error $\epsilon$ is calculated between the target amplitude $A_t$ (in this case, the Max Planck Society Minerva Logo shown in (b)) and the propagated amplitude $A_{t'}$.  The phase mask is then updated based on the error, and the process is iterated until convergence. The resulting phase additions are treated as polarization rotations for this purpose. The calculated polarization mask (7.2 mm) is displayed in (c). The experimentally recorded amplitude distribution (12.3 mm) in the target plane is shown in (d).}
    \label{fig:traditional_holography}
\end{figure*}

\begin{enumerate}
    \item A random phase mask $\phi_s'$ is initialized:
        \begin{equation}
            \phi_{s'} = \textrm{Random phase}
        \end{equation}
    \item The field $U_s$ is then computed from the input amplitude $A_s$ and the random phase mask $\phi_{s'}$:
        \begin{equation}
            U_s = A_s e^{i\phi_{s'}}
        \end{equation}
    \item The field is subsequently propagated over a distance $dz$ to the target plane, modeling free-space diffraction using the ASM:
        \begin{equation}
            U_p = \textrm{Prop}(U_s,dz)
        \end{equation}
    \item The propagated field is then back-propagated to the plane at the origin $z=0$, considering a propagation distance of $-dz$:
        \begin{equation}
            U_{s'} = \textrm{Prop}(U_p,-dz)
        \end{equation}
    \item Next, the phase of this back-propagated field is computed and multiplied by the source amplitude $A_s$ to obtain the updated field $U_s$:
        \begin{equation}
            U_s = A_s e^{i\phi_{s'}}
        \end{equation}
    \item This iterative process is repeated for a maximum number of iterations $i_{\textrm{max}}$.
    \item Upon reaching the maximum number of iterations, the phase mask $\phi_{s'}$ is converted to a polarization rotation angle:
        \begin{equation}
            \alpha_t = \phi_t
        \end{equation}
\end{enumerate}

Following the iterative algorithm described above, we obtain an optimized polarization rotation mask. The efficacy of this mask is subsequently validated experimentally.  The experimentally measured intensity distribution at the target plane is presented in Fig. \ref{fig:traditional_holography}d. For this demonstration, the target image is the Minerva logo of the Max Planck Society, depicted in Fig. \ref{fig:traditional_holography}b. The experimental results demonstrate the capability of the optimized polarization mask to generate the overall structure of the desired amplitude pattern. However, a closer examination of the measured intensity distribution reveals certain limitations. Specifically, the contrast in the reconstructed image is lower than desired. Furthermore, finer features of the target image, particularly the thin curves, are not accurately reproduced. These imperfections underscore the inherent limitations of the traditional Gerchberg-Saxton approach when applied to the optimization of polarization masks for complex intensity patterns.  This observation motivates the exploration of alternative optimization strategies, such as machine learning-based techniques, to further refine the design of polarization masks and achieve improved fidelity in the generated intensity distributions.

\subsection{Holography using Machine Learning Methods}
\label{subsec:machine_learning_methods}

Machine learning optimization methods, including backpropagation, are highly effective in identifying optimal solutions for a given problem. Within the context of holography, machine learning techniques have been successfully applied to optimize phase masks; for instance, in multiplane optimization scenarios. We leverage machine learning optimization techniques to optimize polarization masks for generating arbitrary amplitude patterns at the target plane.

We begin by defining a model, as detailed in Sec. \ref{sec:methods}, comprising a Gaussian source followed by free-space propagation, polarization rotation, and further free-space propagation. Machine learning optimization requires a differentiable model, enabling the calculation of gradients from the error associated with a given input. These gradients facilitate updating the polarization modulation to minimize the error and achieve the closest possible match to the target. As described in Sec. \ref{sec:methods}, the angular spectrum method is employed for propagating the field from the object plane to the target plane. As illustrated in Eq. \ref{eq:asm_composition}, the angular spectrum can be decomposed into a sequence of functions: the field decomposition, forward propagation, and rotation matrix. As each function is differentiable, the entire propagation model is differentiable, enabling the application of machine learning optimizers to optimize the polarization mask. We specifically utilize the Adam optimizer, a variant of the stochastic gradient descent optimizer.

The subsequent component in this process is defining the loss function. The loss function provides a quantitative measure of the discrepancy between the target and the output. Minimizing the error optimizes the calculated mask to generate the desired output upon diffraction. While standard error functions, such as mean square error (MSE) and cross-entropy loss, could be employed, these generic loss functions do not account for the specific requirements of our particular optimization problem. We must consider the following requirements for the specific loss function. These requirements are categorized as follows:

\begin{enumerate}
    \item \textbf{Base Loss Component:} The base loss component quantifies the discrepancy between the target and output patterns without imposing additional constraints. In this context, we employ the mean square error (MSE) loss function. The MSE loss function is defined as the average of the squared differences between the predicted and target images. The base loss function is expressed as:
        \begin{equation}
            L_{\textrm{base}} = \frac{1}{N} \sum_{i=1}^{N} (\textrm{pred}_i - \textrm{target}_i)^2
            \label{eq:mse_loss}
        \end{equation}

    \item \textbf{Dark Region Penalty:} Because the base loss component solely aims to minimize the difference between the target pattern and the output, it might converge to a local minimum where bright regions, corresponding to the intended pattern, are maximized while neglecting the surrounding pixels. However, high contrast in the output is desirable. Consequently, we introduce an additional penalty for a lack of contrast. We define a threshold and identify all pixels in the image where the difference between the predicted output pixels and the target output pixels exceeds this arbitrarily defined threshold:

        \begin{equation}
            M_{\textrm{dark}} = \big(\textrm{pred}_i - \textrm{target}_i\big) > \textrm{threshold}
        \end{equation}

        The dark region penalty component applies an additional penalty to the predicted values within these identified regions. This penalty is weighted by a "dark weight" parameter ($w_{\textrm{dark}}$). The dark region penalty is mathematically expressed as:

        \begin{equation}
            L_{dark} = \frac{1}{N_{\textrm{dark}}} \sum_{i \in M_{dark}} \textrm{pred}_i^2
        \end{equation}

    \item \textbf{Total Loss Calculation:} The total loss is a weighted sum of the base loss and the dark region penalty. The total loss is defined as:
        \begin{equation}
            L_{\textrm{total}} = L_{\textrm{base}} + w_{\textrm{dark}} \cdot L_{\textrm{dark}}
        \end{equation}
\end{enumerate}

\begin{figure*}[!hbt]
    \centering
    \includegraphics[width=0.95\textwidth]{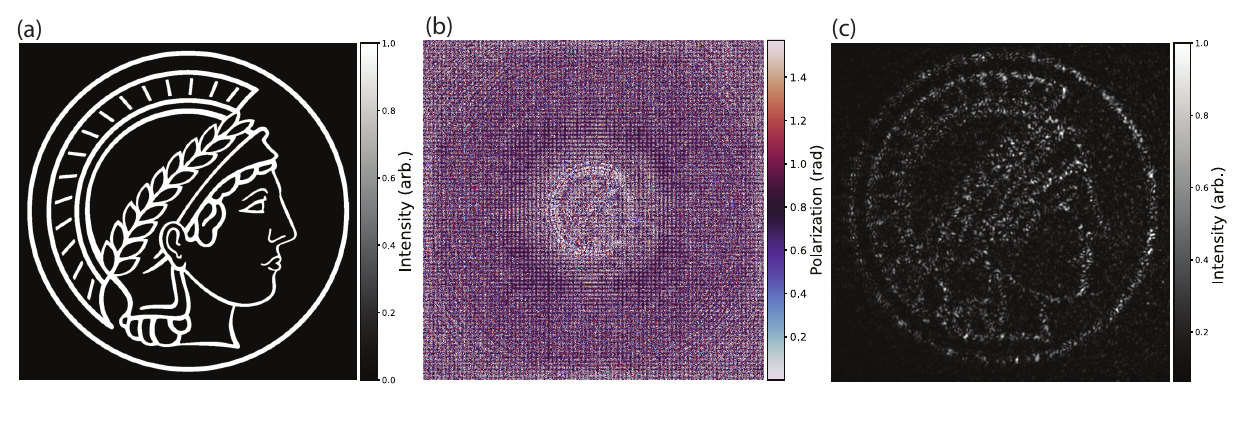}
    \caption{Application of machine learning techniques to optimize a polarization mask to achieve a specific amplitude pattern. The target image, shown in (a), is the Minerva logo of the Max-Planck-Society, consistent with Fig. \ref{fig:traditional_holography}a. The polarization mask (14.4 mm), derived via the machine learning approach, is displayed in (b). The experimentally measured intensity distribution (16 mm) in the target plane is presented in (c).}
    \label{fig:machine_learning_holography_results}
\end{figure*}

This error is then backpropagated through the system to update the polarization rotation angles within the model. This iterative process continues until the error is minimized. The outcomes of this optimization process are presented in Fig. \ref{fig:machine_learning_holography_results}. The same target image as used previously is employed for ease of comparison, as depicted in Fig. \ref{fig:machine_learning_holography_results}a. The polarization mask obtained through the machine learning optimization process is shown in Fig. \ref{fig:machine_learning_holography_results}b. The resultant output, displayed in Fig. \ref{fig:machine_learning_holography_results}c, closely approximates the target image. The output from the optimized mask exhibits enhanced contrast and improved fidelity compared to the results obtained with the adapted Gerchberg-Saxton algorithm.

\subsection{Joint Optimization of Phase and Polarization using Machine Learning}
\label{subsec:joint_optimization}
Beyond optimizing for a target amplitude output, we can also perform joint optimization for both a target amplitude and a target polarization pattern. The model and the machine learning optimization procedure are identical to those detailed in Sec. \ref{subsec:machine_learning_methods}.  The error function is adapted to accommodate the specific requirements of this problem, namely, to account for discrepancies in both amplitude and polarization with respect to the targets.

We introduce a specific loss function, the \texttt{PolAmpLoss}, which is designed to handle both amplitude and polarization targets. The requirements for this specific loss function are:

\begin{enumerate}
    \item \textbf{Amplitude Loss:} The loss accumulated due to the mismatch between the predicted amplitude and the target amplitude is calculated as described in Sec. \ref{subsec:machine_learning_methods}.

    \item \textbf{Polarization Loss:}  When aiming to match the polarization across the wavefront at the output plane with the target polarization, it is crucial to minimize the difference between them.  Therefore, we introduce a custom loss term that explicitly penalizes any deviations, thus enforcing precise predictions.  This composite loss function combines the L1 loss (Equation \ref{eq:l1_loss}) with a conditional exact match penalty (Equation \ref{eq:exact_match_penalty}). The L1 loss ensures overall proximity between predictions and targets, while the exact match penalty strongly penalizes deviations that exceed a predefined threshold ($\epsilon$).

    The L1 loss is given by:
        \begin{equation} \label{eq:l1_loss}
            L_{\text{L1}} = \frac{1}{N} \sum_{i=1}^{N} \left| \text{pred}_i - \text{target}_i \right|
        \end{equation}

    Where $N$ represents the batch size, and $\text{pred}_i$ and $\text{target}_i$ denote the predicted and target values for the $i$-th sample, respectively.

    The exact match penalty (EMP) is $0$ when $\left| \text{pred}_i - \text{target}_i \right| \leq \epsilon$. However, when the difference exceeds $\epsilon$, a quadratic penalty is applied, scaled by a weight factor $w_{\text{exact}}$:
        \begin{equation} \label{eq:exact_match_penalty}
            L_{\text{EMP}} = \frac{1}{N} \sum_{i=1}^{N} w_{\text{exact}} \cdot (\left| \text{pred}_i - \text{target}_i \right| - \epsilon)^2
        \end{equation}

    The loss resulting from the polarization mismatch is the sum of these two components:

        \begin{equation}
            L_{\textrm{polarization}} = L_{\text{L1}} + L_{\text{EMP}}
        \end{equation}

    The exact match penalty is only applied when the absolute difference between the prediction and the target is greater than $\epsilon$.  In these instances, a quadratic penalty, scaled by $w_{\text{exact}}$, is applied, ensuring that larger deviations incur a greater penalty.  This encourages the model to prioritize exact matches when the difference surpasses the defined threshold.

    \item \textbf{Total Loss Calculation:} The total loss is a combination of the base loss, the dark region penalty, and the exact match loss. The total loss is mathematically defined as:
        \begin{equation}
            L_{\textrm{total}} = \alpha \cdot L_{\textrm{intensity}} + (1 - \alpha) \cdot L_{\textrm{polarization}}
        \end{equation}
        where $\alpha$ is a weighting factor that balances the contributions of the intensity and polarization losses.
\end{enumerate}

\begin{figure*}[!hbt]
    \centering
    \includegraphics[width=0.95\textwidth]{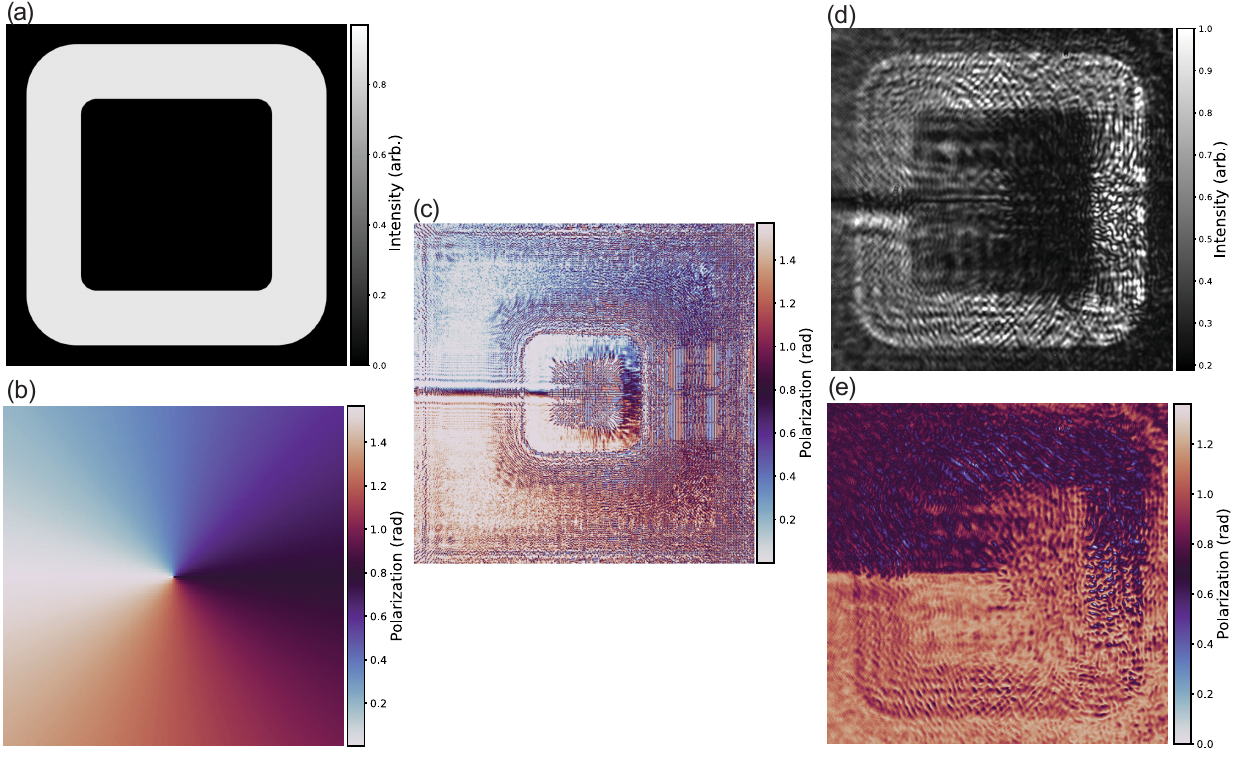}
    \caption{Joint optimization of phase and polarization employing machine learning techniques.  The target image, a rounded square pattern, is shown in (a).  The target polarization, depicted in (b), features a polarization angle that sweeps from $0$ to $2\pi$ radians from the center.  The polarization mask (14.4 mm) obtained through the joint optimization process is displayed in (c).  The resulting intensity distribution (10.6mm), recorded using a camera in the target plane, is presented in (d), and the calculated polarization from experimental measurements is shown in (e).}
    \label{fig:machine_learning_holography_results_joint_optimization_simple}
\end{figure*}

By utilizing the loss function in conjunction with the ADAM optimizer described previously, we optimize for a polarization modulation which, upon beam propagation, yields an amplitude pattern and a polarization pattern that match the target amplitude and polarization patterns. Because the amplitude and polarization in the target plane are coupled, and we have limited degrees of freedom (in the form of a polarization mask) to achieve the targeted change, we initially restrict ourselves to a simple target.  We first consider a rounded square amplitude target, shown in Fig. \ref{fig:machine_learning_holography_results_joint_optimization_simple}a, and a polarization pattern, shown in Fig. \ref{fig:machine_learning_holography_results_joint_optimization_simple}b, that sweeps from $0$ to $2\pi$ radians from the center. After executing the optimization routine, we obtain a polarization mask, as shown in Fig. \ref{fig:machine_learning_holography_results_joint_optimization_simple}c. As can be observed in Fig. \ref{fig:machine_learning_holography_results_joint_optimization_simple}d, the measured output corresponds to the target, with some additional diffraction substructure that is difficult to eliminate entirely due to the thickness of the bright lines in the target.

We measure the polarization pattern at the output plane by positioning a polarizer in the beam path after the SLM and scanning over a polarization range between $0$ radians and $\pi$ radians, resulting in two complete cycles of change. From this image series, we can reconstruct the scan, as shown in SUPP. VIDEO 1. Moreover, we can calculate the polarization at each point of the detected image using one of the images, according to Malus' law. However, we repeat this calculation for four distinct images to improve statistical robustness. The computed polarization is presented in Fig. \ref{fig:machine_learning_holography_results_joint_optimization_simple}e. The polarization pattern closely resembles the target polarization pattern, with the polarization angle sweeping from $0$ to $2\pi$ radians from the center. This underscores the advantage of optimizing the polarization mask, providing an additional degree of freedom when combined with phase modulation to attain complex amplitude and polarization patterns at the target plane.

\begin{figure*}[tp]
    \centering
    \includegraphics[width=0.95\textwidth]{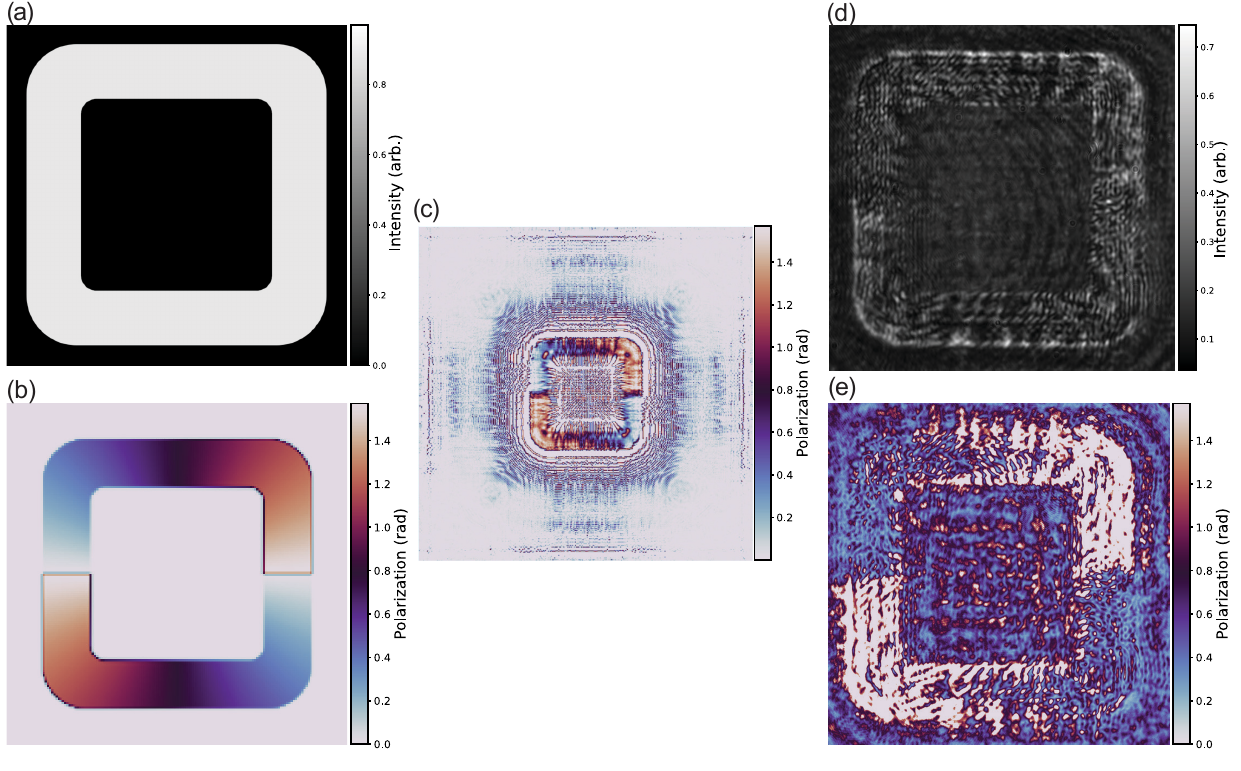}
    \caption{Optimization of complex polarization patterns using machine learning techniques.  The target image, depicted in (a), comprises a square amplitude distribution.  The target polarization, shown in (b), features a structured sweep of polarization angles ranging from $0$ to $\pi/2$, repeated twice along the square loop. The polarization mask (14.4 mm), derived through the joint optimization process, is presented in (c).  The experimentally obtained amplitude distribution (10.6 mm) in the target plane is shown in (d), and the calculated polarization from experimental measurements is illustrated in (e).}
    \label{fig:machine_learning_holography_results_joint_optimization_complex}
\end{figure*}

Given that jointly optimizing for polarization and amplitude involves a trade-off between the two, we sought to further investigate the capability of this technique to jointly optimize for a more complex polarization target.  We maintain the same target amplitude pattern as before, using a rounded square as shown in Fig. \ref{fig:machine_learning_holography_results_joint_optimization_complex}a, but we modify the target polarization pattern to a structured sweep of polarization angles ranging from $0$ to $\pi/2$, repeated twice along the square loop, as shown in Fig. \ref{fig:machine_learning_holography_results_joint_optimization_complex}b. The polarization mask obtained through the joint optimization procedure is presented in Fig. \ref{fig:machine_learning_holography_results_joint_optimization_complex}c. Although the experimentally obtained amplitude output exhibits lower contrast than in the simpler case, as shown in Fig. \ref{fig:machine_learning_holography_results_joint_optimization_complex}d, it is evident that the polarization mask successfully generates the desired amplitude pattern. We adhere to the same procedure as previously described to optimize the polarization mask and measure the polarization pattern on the target plane. The video of the polarization scan is presented in SUPP. VIDEO 2. The calculated polarization pattern is illustrated in Fig. \ref{fig:machine_learning_holography_results_joint_optimization_complex}e. As observed, the polarization pattern closely matches the target polarization pattern, thereby demonstrating the capability of this technique to jointly optimize phase and polarization masks to achieve intricate amplitude and polarization patterns at the target plane, utilizing only the degrees of freedom provided by a spatially patterned polarization mask.

\section{Discussion}
\label{sec:discussion}

In this study, we have demonstrated that spatial variations in polarization across a beam profile,  can be controlled to precisely control the amplitude and polarization state of the propagated wavefront at a target location downstream.  This expands upon the traditional understanding of diffraction, which typically focuses on phase or amplitude variations.  We show that a wave with uniform phase, but a suitable polarization distribution across its wavefront can diffract into complex desired intensity amplitude patterns. By combining free-space propagation with continuous polarization masks, we have thus demonstrated the potential to perform a variety of beam-shaping and manipulation operations.  This concept was experimentally validated by transforming a Gaussian beam into a focused beam akin to Bessel beam through spatial polarization modulation.

However, as we have shown, calculating the required polarization mask for arbitrary target intensity patterns is non-trivial. To address this, we adapted the well-known Gerchberg-Saxton algorithm, traditionally used in phase retrieval holography, to the domain of polarization modulation. While this adapted algorithm successfully generated a polarization mask to produce the Minerva logo, a machine-learning-based optimization technique that we developed yielded superior results in terms of fidelity and accuracy. Furthermore, the machine learning approach offers greater flexibility.  We demonstrated this by extending it to achieve simultaneous and independent control of both amplitude and polarization at the target plane, using only a single polarization modulation step.  This capability opens up possibilities for manipulating light in a more comprehensive way than previously achievable.

The results presented here have several significant implications. Firstly, they establish spatial polarization modulation as a powerful tool for wavefront shaping, offering an alternative to conventional phase and amplitude modulation techniques. This opens up the potential to combine polarization modulation with phase modulation to obtain complete control over amplitude, polarization, and phase at the output plane. Moreover, this approach could enable a spatially evolving amplitude, phase, and polarization as the wavefront propagates, opening up further applications in microscopy and lithography.  Additionally, polarization-based beam shaping may be particularly advantageous in applications where preservation of the beam's phase is crucial, such as in certain interferometric setups.

\section{Data Availability}
The data and code that support the findings of this study are available from the corresponding author upon reasonable request.

\section{Acknowledgements}

The authors express their gratitude to Dr. B. Schölkopf, Dr. V. Volchkov, and Mr. L. Schlieder for their insightful feedback and comments. 

\section{References}
\bibliography{references}

\setcounter{page}{1}
\clearpage
\onecolumngrid
\clearpage
\newpage
\newpage
\clearpage  
\setcounter{figure}{0}
\setcounter{table}{0}
\setcounter{section}{0}
\setcounter{equation}{0}
\setcounter{page}{1}
\setcounter{footnote}{0}

\renewcommand{\theequation}{S\arabic{equation}}
\renewcommand{\thefigure}{S\arabic{figure}}
\renewcommand{\thetable}{S\arabic{table}}
\renewcommand{\thesection}{S\arabic{section}}
\renewcommand{\thesubsection}{S\arabic{section}.\arabic{subsection}}
\renewcommand{\thesubsubsection}{S\arabic{section}.\arabic{subsection}.\arabic{subsubsection}}

\begin{center}
  {\Large\bfseries Supplementary Information for:\\[0.5em]
  Engineering wavefronts with machine learned structured polarization}\\[1em] 
  Sai Nikhilesh Murty Kottapalli\textsuperscript{1,2},
  Alexander Song\textsuperscript{1,2},
  Peer Fischer\textsuperscript{1,2,3,4}\\[0.5em]
  \small
  \textsuperscript{1}Max Planck Institute for Medical Research, 69120 Heidelberg, Germany\\
  \textsuperscript{2}Institute for Molecular Systems Engineering and Advanced Materials,\\
  Universität Heidelberg, 69120 Heidelberg, Germany\\
\textsuperscript{3}Center for Nanomedicine, Institute for Basic Science (IBS), Seoul, Republic of Korea\\
    \textsuperscript{4}Department of Nano Biomedical Engineering (NanoBME), Advanced Science Institute, Yonsei University, Seoul, Republic of Korea\\
\end{center}

\section{Contribution of Geometric Phase}
\label{si:geometric_phase}

The wavefront shaping technique described in this work relies on spatially varying polarization rotation imparted by a spatial light modulator (SLM).  Since we are considering only linearly polarized states in this instance, we are constrained to the equator of the Poincaré sphere. For a given input field

\begin{equation}
    E = \begin{pmatrix}
    E_x \\
    E_y
    \end{pmatrix}
    \label{seq:jones_vector}
\end{equation}

with two orthogonal linear components, $E_x$ and $E_y$, the twisted nematic SLM utilizes birefringence to induce a polarization rotation, $\theta$, that can be described using the following Jones matrix:

\begin{equation}
    M = \begin{pmatrix}
    \cos(\theta) & -\sin(\theta) \\
    \sin(\theta) & \cos(\theta)
    \end{pmatrix}
    \label{seq:jones_rotation_matrix}
\end{equation}

As the field rotates around the equator of the Poincaré sphere, it acquires a geometric phase, which can be calculated from the solid angle traced by the loop. For a rotation operation completing a full $2\pi$ rotation of polarization, the geometric phase acquired is:

\begin{equation}
    \Phi_{\textrm{GP}} = -\frac{\Omega}{2}
\end{equation}

As previously mentioned, we consider only linear polarization transformations here.  Consequently, we obtain a maximum geometric phase gain of $-\pi$ radians.  However, we are constrained by the SLM used to operate within a polarization rotation range of $0$ to $\pi/2$. Thus, we need to consider the geometric phase gained due to non-closed paths on the Poincaré sphere. For any non-closed path traced by the SLM operation, the geometric phase can be calculated as half of the solid angle between the initial and final states. In the experimental case under discussion, this corresponds to a maximum geometric phase gain of $-\pi/4$ radians.

\begin{figure}[h]
    \centering
    \includegraphics[width=0.95\textwidth]{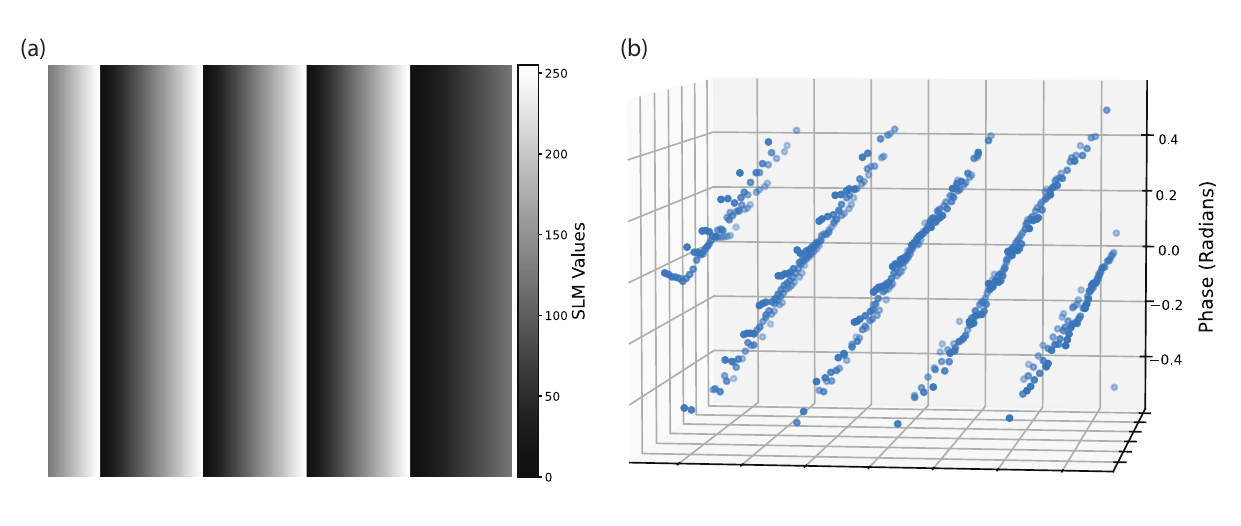}
    \caption{Geometric phase accumulation due to polarization rotation. (a) Polarization sweep pattern projected onto the SLM. (b) Measured phase shift due to polarization rotation.}
    \label{sfig:geometric_phase}
\end{figure}

We experimentally verify this for the optical system shown in FIG. \ref{fig:schematic} by projecting a polarization-varying pattern onto the SLM and imaging it onto a phase detector, in this case a Shack-Hartmann sensor. The results of such a measurement are displayed in SUPP. FIG. \ref{sfig:geometric_phase}. We project a sweep pattern onto the SLM with projected values ranging from 0 to 255, corresponding to a polarization rotation of 0 to $0.5\pi$, as shown in SUPP. FIG. \ref{sfig:geometric_phase}a.  The phase shift measured by the Shack-Hartmann sensor is shown in SUPP. FIG. \ref{sfig:geometric_phase}b. As can be observed from the figure, the measured and calculated geometric phase accumulation closely match each other.

\begin{figure}[tp]
    \centering
    \includegraphics[width=0.95\textwidth]{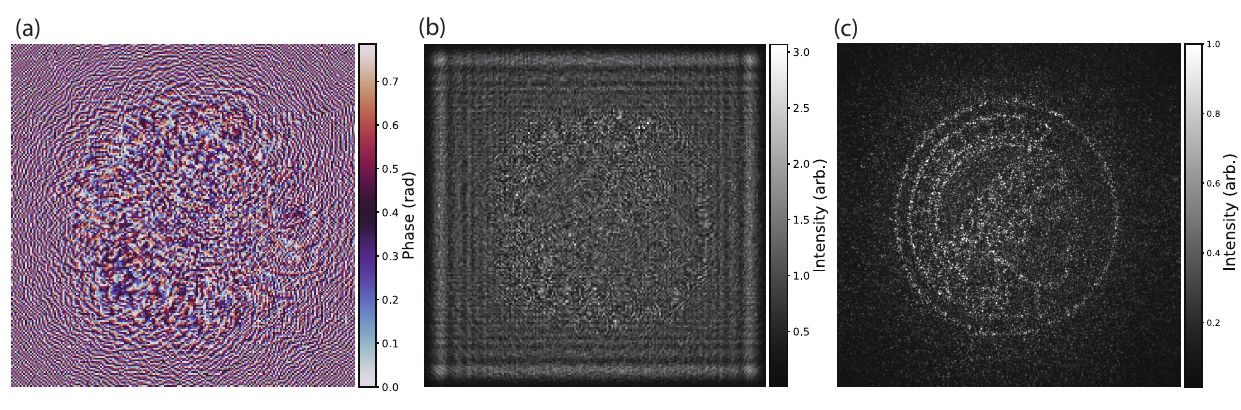}
    \caption{Effect of geometric phase accumulation on the final intensity distribution. (a) Phase sweep pattern projected onto the SLM. (b) Intensity distribution at the target plane with phase modulation. (c) Intensity distribution in the target plane with polarization modulation.}
    \label{sfig:geometric_phase_sim}
\end{figure}

We can subsequently analyze how a change in the phase of the propagating wavefront, by a magnitude $\Delta\phi$, affects the amplitude distribution at the target plane.  We employ the angular spectrum method, as described in Sec. \ref{sec:methods}, to simulate the propagation of a Gaussian beam with a phase modulation instead of the polarization modulation.  We use the polarization mask obtained in Sec. \ref{subsec:traditional_methods} as an example and scale the corresponding phase modulation to be between $0$ and $\pi/4$, as shown in SUPP FIG. \ref{sfig:geometric_phase_sim}a.  With this scaled phase modulation, we obtain an intensity distribution on the target plane, as shown in SUPP FIG. \ref{sfig:geometric_phase_sim}b.  Comparing this with the output obtained by considering the polarization mask shown in SUPP FIG. \ref{sfig:geometric_phase_sim}c, it is evident that the polarization mask output is closer to the target intensity distribution than what could be achieved with the geometric phase contribution. This demonstrates that the effect of geometric phase accumulation on the final intensity distribution can be safely neglected.

\section{Focusing using Polarization Masks}
\label{si:focussing}
\begin{figure}[!hbt]
    \centering
    \includegraphics[width=0.95\textwidth]{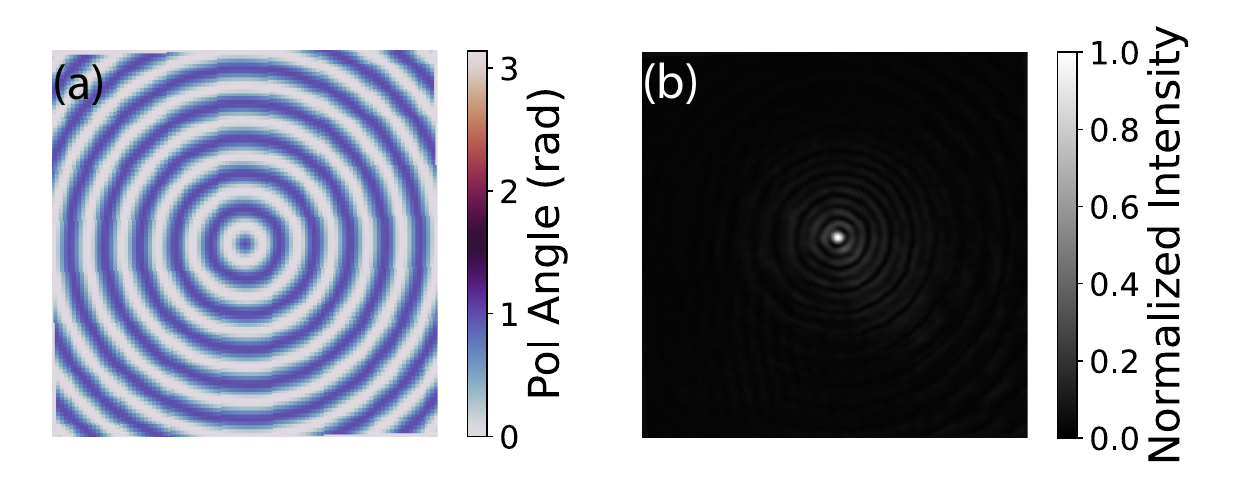}
    \caption{Generation of a Bessel beam using a polarization mask. (a) The polarization mask applied to the spatial light modulator (SLM). (b) The resultant amplitude distribution on the target plane, exhibiting the characteristic non-diffracting nature of the Bessel beam, achieved in this case with a polarization mask.}
    \label{sfig:bessel_beam}
\end{figure}

We first attempt to implement the simple case of focusing a Gaussian beam using only polarization modulation. A commonly used beam for focusing is the Bessel beam. We design a polarization rotation mask, as shown in Fig. \ref{sfig:bessel_beam}a, which we project onto the SLM. The mask is designed to rotate the polarization of the beam in such a way that it generates a focused spot on the target plane, akin to a Bessel beam. The resultant measured intensity pattern is shown in SUPP. FIG. \ref{sfig:bessel_beam}b. The intensity distribution exhibits a central focused spot with concentric rings surrounding it. Each of the rings contains equal energy, leading to a Bessel-like intensity distribution. However, it should be noted that this is not a true, non-diffracting Bessel beam.

\end{document}